\documentclass[dvips]{article}

\usepackage[utf8]{inputenc}

\usepackage{textcomp}
\usepackage{booktabs}

\usepackage{icrc2011}

\title{First results of the CROME experiment}

\newcommand{\etal}{\MakeLowercase{\textit{et al. }}} 
\shorttitle{R.~\v{S}m\'{\i}da \etal First results of the CROME experiment}

\authors{
R.~\v{S}m\'{\i}da$^{1}$, H.~Bl\"{u}mer$^{1}$, R.~Engel$^{1}$, A.~Haungs$^{1}$, T.~Huege$^{1}$,
K.-H.~Kampert$^{2}$, H.~Klages$^{1}$, M.~Kleifges$^{1}$, O.~Kr\"{o}mer$^{1}$, S.~Mathys$^{2}$,
J.~Rautenberg$^{2}$, M.~Riegel$^{1}$, M.~Roth$^{1}$, F.~Salamida$^{3}$, H.~Schieler$^{1}$, J.~Stasielak$^{4}$,
M.~Unger$^{1}$, M.~Weber$^{1}$, F.~Werner$^{1}$, H.~Wilczy\'{n}ski$^{4}$, J.~Wochele$^{1}$
}
\afiliations{
$^1$Karlsruhe Institute of Technology (KIT), Karlsruhe, Germany\\
$^2$Bergische Universit\"{a}t Wuppertal, Wuppertal, Germany\\
$^3$Universit\`{a} dell'Aquila and INFN, L'Aquila, Italy\\
$^4$Institute of Nuclear Physics PAN, Krakow, Poland
}
\email{radomir.smida@kit.edu}

\abstract{
It is expected that a radio signal in the microwave range
is produced in the atmosphere due to molecular bremsstrahlung
initiated by extensive air showers. The CROME (Cosmic-Ray
Observation via Microwave Emission) experiment was built to search
for this microwave signal. Radiation from the atmosphere is
monitored in the extended C~band (3.4--4.2\,GHz) in coincidence
with showers detected by the KASCADE-Grande experiment. The detector
setup consists of several parabolic antennas and fast read-out
electronics. The sensitivity of the detector has been measured with
different methods. First results after half a year of data taking
are presented.
}


\begin{document}
\maketitle

\section{Introduction}

Low-energy electrons in extensive air showers (EAS) are expected to
produce radio radiation at microwave frequencies. Beam measurements of
this process provided a first estimate of the intensity of the microwave
signal~\cite{gorham}. The unpolarized and isotropic microwave signal
has been attributed to molecular bremsstrahlung radiation due
to the interaction of electrons with neutral air molecules. If this
radiation can be measured by standard GHz-radio instruments, it will
provide a new and promising way of the observation of ultra-high
energy cosmic rays (UHECRs)~\cite{gorham}.

Microwave radiation (1--10\,GHz) from EAS could provide information
about the longitudinal development of EAS and, hence, the energy and
chemical composition of the primary cosmic-ray particles.  The
advantages of a microwave measurement with respect to an observation
of fluorescence light (see e.g.~\cite{bluemer}) are the 100\% on-time
and the very small atmospheric attenuation of the signal even in the
presence of clouds. The two frequency bands C (3.7--4.2\,GHz) and Ku
(10.7--12.7\,GHz) for satellite communication are particularly
attractive as they are characterized by very low natural background radiation
and negligible human made radio frequency interference~\cite{erc}. Moreover,
commercial equipment for low-noise receivers for the C and Ku bands is commonly
available.

Several projects aiming at measuring microwave emission from EAS have
been started~\cite{privitera} but a signal detection has not yet been
reported. In this work we discuss the status and calibration of the
CROME (Cosmic Ray Observation via Microwave Emission) detector that
has been set up within the KASCADE-Grande (KG) air shower array near
Karlsruhe, Germany~\cite{grande}.

\section{Detector setup}

\begin{figure}[!t]
\vspace{5mm}
\centering
\includegraphics{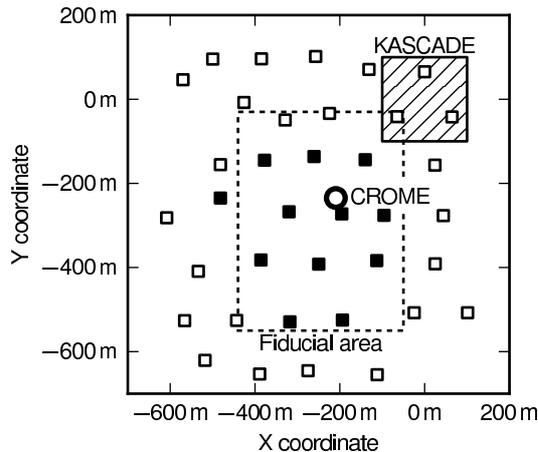}
\caption{ Location of CROME antennas in the KASCADE-Grande array. Full
  squares indicate the scintillator stations that provide the trigger
  for CROME. The information from the smaller but much denser KASCADE
  array~\cite{kascade} (upper right corner) with separate electron and
  muon counters is used for estimating the number of muons.  }
\label{kg-array}
\end{figure}

The CROME experiment aims at the detection of microwave signals from
air showers as expected from molecular bremsstrahlung and Cherenkov
emission. Commercially available GHz antennas are used to continuously monitor the air above
the KASCADE-Grande array. Using a shower trigger provided by
KASCADE-Grande~\ref{kg-array}, the measured GHz signal is stored for all high-energy
showers detected with the scintillator array. 

The antenna setup consists of several microwave receivers for the
frequency ranges 1--1.8, 3.4--4.2 (C band), and 10.7--11.7\,GHz (low
Ku band). While the 1--1.8\,GHz antenna is a scientific instrument
originally built for observing the 21\,cm hydrogen line, the other
antennas are commercially available parabolic reflectors equipped with
low noise block-downconverters (LNBs). All antennas are oriented
vertically upwards.

\begin{figure}[!t]
\vspace{5mm}
\centering
\includegraphics[width=2.8in]{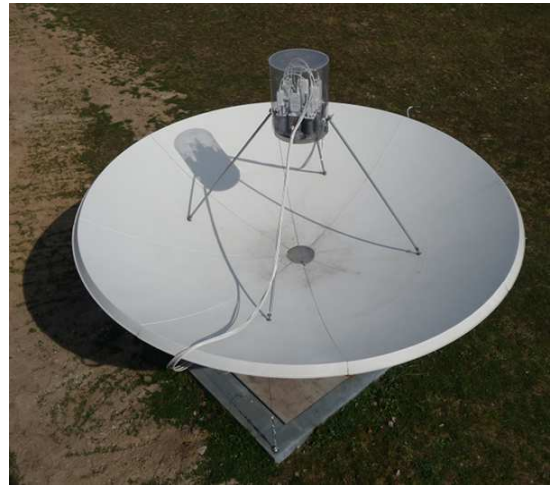}
\caption{
Photo of segmented parabolic dish (Prodelin 1344) of 335\,cm diameter
and 119\,cm focal length. The camera of 9 linearly polarized C~band
receivers, each consisting of a feed matched to the antenna size and a
Norsat 8215F LNB, is supported by four struts.
}
\label{crome-dish}
\end{figure}

The current setup of CROME consists of a 2.3\,m dish with one receiver
for 1--1.8\,GHz, two 3.4\,m dishes each with a camera with 9 C~band
receivers (see Figure~\ref{crome-dish}) and a small 0.9\,m dish with
one receiver for the Ku band. The field of view of a single C band
receiver in the 3.4\,m antenna corresponds to an opening angle of
$\sim$ 1.6$^{\circ}$ for a drop in the signal by 3\,dB (for more
details see Section~\ref{section-calib}).

The layout of the setup is shown in Figure~\ref{kg-array}. Twelve
scintillator stations in the center of the KASCADE-Grande array,
indicated by full squares in the figure, provides the trigger with a
rate of a few hundred events per day.  Only showers with the core
position reconstructed inside the area marked by the dashed lines in
Fig.~\ref{kg-array}, corresponding to $2.0\times10^{5}$\,{\rm m}$^2$,
are used in the data analysis.

A fast logarithmic power detector (Analog Devices AD8318) is used to
measure the envelope of the antenna signals.  Measurements show that
the exponential time constant of the whole chain of electronics
together with the LNB is $\sim$4\,ns (i.e.{} $\sim$9\,ns for 10-to-90\%
risetime).  The measured signal is readout by five 4-channel
digitizers (PicoScope 6402 and 6403) with a sampling time of 0.8\,ns
and 8-bit dynamic range. All channels are readout in a time window of
10\,\textmu before and after the trigger delivered by KASCADE-Grande.

Particular attention has been paid to the time synchronization between
the CROME and KASCADE-Grande DAQs since the radio signal is expected
to be as short as 20\,ns for nearly vertical showers. A GPS clock
identical to the one of KASCADE-Grande is used (Meinberg 167 lcd) and,
in addition, the trigger time from one scintillator station of
KASCADE-Grande is stored for each event. The reconstructed arrival
direction, core position, and energy of the showers recorded by the
KASCADE-Grande array are used in the further analysis.

\section{Expected event rate}
\label{section-expected}

The calculation of the number of showers measurable by the CROME antennas is based on a detailed simulation
of air showers detected by KASCADE-Grande together with the estimated microwave signal according to~\cite{gorham}.

KASCADE-Grande is optimized to detect air showers in the range from
$10^{15.5}$ to $10^{18}$\,eV. On average, one shower above
$10^{17}$\,eV per day is measured by KASCADE-Grande and successfully
reconstructed. Details can be found in~\cite{grande}. The energy range
of KASCADE-Grande includes also the energy of $3.4\times 10^{17}$\,eV
for which Gorham et al.~\cite{gorham} made the accelerator
measurement.

Lacking a detailed microscopic model for microwave emission in air
showers, we assume that a fixed fraction of the energy $E_{\rm dep}$ deposited by a
shower in the atmosphere is radiated off in the microwave frequency
band $\Delta \nu$. This fraction, in the following called microwave
yield $Y_{\rm MW}$,
is determined by comparing the simulation results for air showers with the expected
signal given in~\cite{gorham}
\begin{equation}
Y_{\rm MW} = \frac{1}{\Delta \nu} \frac{E_{\rm MW}}{E_{\rm dep}} \approx
1.2\times10^{-18}\,{\rm Hz}^{-1}.
\end{equation}
Using a 3-d simulation of the showers~\cite{gora} with iron nuclei as
primary particle and neglecting attenuation in the atmosphere, which
is lower than 0.01\,dB/km below 10\,GHz, we get $\sim$\,2 showers per
9 C~band receivers in a 335\,cm dish per month with a microwave signal
above the minimum detectable flux. The minimum detectable flux is
given by $k_{\rm B} T_{\rm sys}/A_{\rm eff}/\sqrt{\tau \Delta\nu}$, where
$k_{\rm B}$ is
the Boltzmann constant, $T_{\rm sys}$=80\,K the system temperature,
$A_{\rm eff}$=6.4\,m$^2$ the effective area of a dish,
$\Delta\nu$=600\,MHz the bandwidth, and $\tau$=10\,ns the integration
time. It should be noted that the parameters of the calculation were
chosen to obtain a robust lower limit to the expected rate for the
given microwave yield.

\section{Calibration and radiation pattern}
\label{section-calib}

A end-to-end system temperature of 60\,K is estimated for the C~band
system by comparing the measurement for clear sky with the measurement
with a microwave absorber (radiating as black body at ambient
temperature) placed in front of the camera.

\begin{figure}[!t]
\vspace{5mm}
\centering
\includegraphics[width=2.0in]{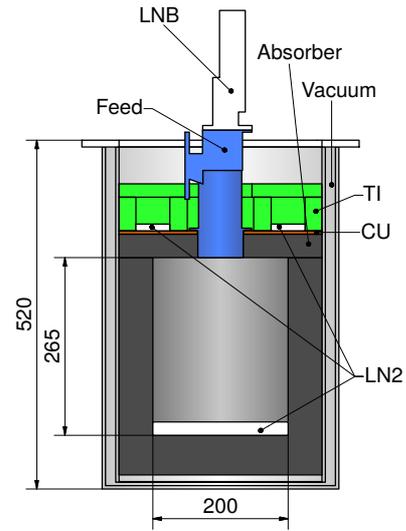}
\caption{ Cryostat built for the calibration of feeds with LNBs. The
  copper plate (CU) supports the feed, closes the inner part of the
  cryostat and holds temperature insulation material (TI). Liquid
  nitrogen (LN2) is filled in the upper part above the copper lid and
  also inside the cryostat which is fully covered by microwave
  absorber material.  }
\label{cryostat}
\end{figure}

The calibration of the individual receivers (LNBs with feeds) has been
measured with a microwave absorber at room and liquid nitrogen
temperature of 77\,K. The setup is shown in Figure~\ref{cryostat}. The
flat 5\,cm thick microwave absorber is placed along a steel wall, in
the bottom and also upper part of a cryostat which is covered by a
copper lid.  The feed can be mounted in stable position and its
entrance is below the copper plate. The liquid nitrogen has been
filled inside the cryostat and also in the upper part above the copper
lid. The temperature has been kept uniform and stable in the whole
inner part of the cryostat. The difference in the measured voltage
corresponds to the system temperature ranging from 40 up to 50\,K for
the tested LNBs.

The antennas do not have an astronomical mount and are kept in stable
positions. Therefore a flying radio source has been developed to study the
sensitivity pattern of the antennas. The radio transmitter is mounted to
a remote-controlled copter together with differential GPS for the
precise measurement of the position. A two-element Yagi antenna with a
one-sided main lobe (about 4.1\,dBi) and high backward attenuation
(-10\,dB) to avoid reflections of the copter is used as the
radio transmitter.  The Yagi antenna consists of a half-wavelength dipole
with a radiation coupled reflector in quarter-wavelength distance.
The maximum power generated by the stabilized transmitter is 8\,dBm (6\,mW) and
covers the frequency range between 2970 and 3950\,MHz. Six operating
modes with different modulation patterns are implemented, which may be
triggered by an external source or internal 3\,Hz clock.  The antenna
provides a wide main lobe with -3\,dB beamwidth of 110$^{\circ}$ on
average. Moreover, a logarithmic-periodic dipole antenna is being
prepared for wideband sweep operating modes.

\begin{figure}[!t]
\vspace{5mm}
\centering
\includegraphics{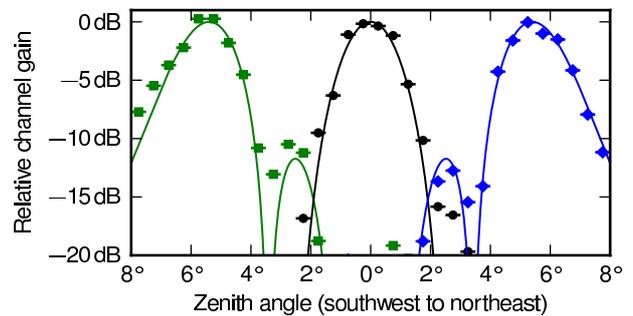}
\caption{ Two-dimensional radiation pattern of the central and two
  off-center receivers measured by the airborne radio source.  Points
  indicate measured values and curves show simulated far field
  radiation patterns. The gain pattern has been normalized separately
  for each feed.  }
\label{gain-pattern}
\end{figure}

Preliminary results in the transition zone between the near and far
field of the CROME antennas 
allow us to determine the radiation pattern of the whole
system. Relative radiation patterns for the central and two off-center
receivers are shown in Figure~\ref{gain-pattern}.  The obtained
results are consistent with simulations for the far field zone
obtained with the software package GRASP~\cite{grasp}.  The main lobes
are clearly visible and the -3\,dB beam-widths are less than
2$^{\circ}$ for all channels. An effect of aberration has been
measured for off-center receivers. The first side lobe is down by more
than 10\,dB relative to the peak of the main beam and is pointed
towards the main lobe of the central receiver.

\section{Performance and data analysis}

CROME is taking data since September 2010. The initial C~band
configuration has been enlarged from a single antenna with 4~receivers
to two antennas with 18 receivers. The second antenna has been
installed in April 2011. A few important changes have been made
during the measurement apart from the installation of new receivers and
read-out electronics. The first was an implementation of inline
filters for the measurement of the average power (fluctuations of
frequencies less than 100\,kHz) together with fast fluctuations. Later
on high pass filters (MiniCircuits VHF-1200) have been installed in
the electronic chain to suppress signals from airplane altimeter radars.

The uptime and number of reconstructed events crossing the field of
view of at least one C~band receiver is shown in
Table~\ref{table-performance} for the given a detector configuration. The
uptime of the detector has been affected in particular by the downtime
of KASCADE-Grande and then by detector upgrades, calibration or test
measurements. Total uptime (i.e.{} period for which CROME and KASCADE-Grande
provided data and full reconstruction of KG events was possible)
equals to more than 170 days with 8.6 receivers taking data on
average.

\begin{figure}[!t]
\vspace{5mm}
\centering
\includegraphics[width=3.0in]{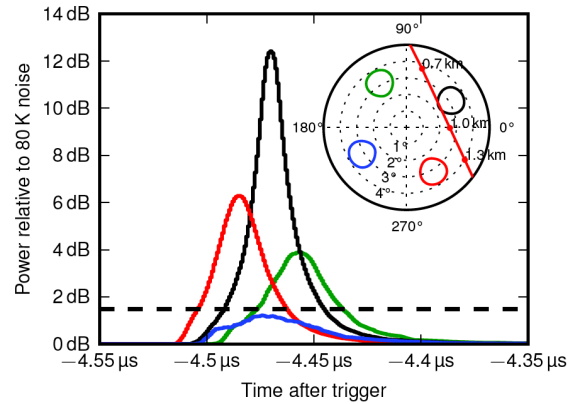}
\caption{ Simulated power flux of the most energetic event crossing
  the field of view of the 4-receiver camera.  The dashed line
  corresponds to the minimum detectable flux (see
  Section~\ref{section-expected}).  The fields of view of the
  receivers are shown in an inset along with the shower track and its
  altitude above the antenna.  }
\label{event-sim}
\end{figure}

In total 79 events with energy higher than $5\times10^{16}$\,eV were
detected by KASCADE-Grande that pass the quality criteria for
reconstruction and have crossed a field of view of at least one
receiver in the C~band setup.  The most energetic event was measured
on September 24th, 2010.  Its energy is $7.9\times10^{17}$~eV with a
zenith angle of $10.5^{\circ}$ and a distance between the shower core
to the antenna of 159\,m. The simulated signal for this event is shown
in Figure~\ref{event-sim}. The highest expected signal is about 4\,dB
above noise fluctuations and the time width of the signal is about
20\,ns.

It is worth to mention that we might detect also a Cherenkov signal
from almost vertical events with reconstructed shower-core distances
less than 100\,m from the antenna (see e.g.~\cite{kalmykov}).

\begin{table}[t]
\begin{center}
  \caption{ Performance of the CROME C~band setup till May 10th,
    2011. The columns show the start dates of measurement with a given
    setup, the number of receivers used for the measurement, the
    uptime in days (percent) and the number of well reconstructed
    events above $5\times10^{16}$\,eV passing through field of view.
  }
\label{table-performance}
\begin{tabular}{ccrlc}
\\
\toprule
Date & Receivers & \multicolumn{2}{c}{Uptime} & Events \\
\midrule
14/09/2010 &  4 &  33.5\,d &  (48\%) &  10 \\
19/11/2010 &  8 & 106.7\,d &  (82\%) &  49 \\
01/04/2011 & 15 &  26.8\,d &  (83\%) &  17 \\
04/05/2011 & 18 &   4.9\,d & (100\%) &   3 \\
\addlinespace
all        & -- & 171.8\,d &  (73\%) &  79 \\
\bottomrule
\end{tabular}
\end{center}
\end{table}

\section{Conclusion}

The first results obtained with the CROME antenna array have been
presented. The detector has shown stable performance over more than
half a year complemented by several improvements in the setup or test
measurements. We have built an airborne GHz transmitter which has
been successfully used for mapping the sensitivity pattern of the
antennas and will be used also for absolute calibration. Also the
properties of the receivers have been measured with a dedicated
test setup. Many extensive air showers initiated by primary particles
with an energy above $5\times10^{16}$\,eV have crossed the field of
view of at least one installed receiver. The analysis of the measured data
is in progress, with the candidate events being studied in detail.

{\bf Acknowledgements}\\
It is our pleasure to thank our colleagues from the Pierre Auger
and KASCADE-Grande Collaborations for many stimulating discussions. 
This work was partially supported by the 
Polish Ministry of Science and Higher Education
under grant No.~NN~202 2072 38
and in Germany by the DAAD, project ID 50725595.

\clearpage
\end{document}